\documentclass[fleqn,usenatbib]{mnras}

\usepackage{newtxtext,newtxmath}

\usepackage[T1]{fontenc}

\DeclareRobustCommand{\VAN}[3]{#2}
\let\VANthebibliography\thebibliography
\def\thebibliography{\DeclareRobustCommand{\VAN}[3]{##3}\VANthebibliography}

\usepackage{graphicx}	
\usepackage{amsmath}	
\usepackage{float}

\title[The first outburst of the AM~CVn SDSS~J113732+405458]{The outburst of a 60 min AM CVn reveals peculiar color evolution: implications for outbursts in long period double white dwarfs}

\author[L.E. Rivera Sandoval]{
L. E. Rivera Sandoval$^{1,2}$\thanks{E-mail: lriveras@ualberta.ca}, T. J. Maccarone$^{2}$, Y. Cavecchi$^3$, C. Britt$^4$ and D. Zurek$^5$\\
$^{1}$University of Alberta, Department of Physics, CCIS 4-183, T6G 2E1, Edmonton, AB, Canada\\
$^{2}$Texas Tech University, Department of Physics \& Astronomy, Box 41051, Lubbock, TX 79409, USA\\
$^{3}$Universidad Nacional Aut\'onoma de M\'exico, Instituto de Astronom\'ia, Ciudad Universitaria, CDMX 04510, Mexico\\
$^{4}$Space Telescope Science Institute, 3700 San Martin Dr., Baltimore, MD 21218, USA\\
$^{5}$American Museum of Natural History, Department of Astrophysics, Central Park West 79th Street, New York, NY 10024, USA\\
}

\date{Accepted XXX. Received YYY; in original form ZZZ}

\pubyear{2021}

\begin{document}
\label{firstpage}
\pagerange{\pageref{firstpage}--\pageref{lastpage}}
\maketitle

\begin{abstract}
We report on multi-wavelength observations during quiescence and of the first detected outburst of the $\approx60$~min orbital period AM~CVn
SDSS~J113732+405458. Using X-ray and UV observations we determined an
upper limit duration of the event of about one year. The
amplitude of the outburst was remarkably small, of around one
magnitude in $r$ and 0.5 magnitudes in $g$. We have also investigated
the color variations of SDSS~J113732+405458 and other long period
AM~CVns in outbursts and identified a track on the color-magnitude
diagram that is not compatible with the predictions of the disk
instability model, suggesting that some outbursts in long period
AM~CVns are caused by enhanced mass-transfer. To our knowledge, these
are the first studies of the color evolution in AM~CVns.
During quiescence we measured an X-ray luminosity for
SDSS~J113732+405458 of $\approx3\times 10^{29}$ erg/s in the 0.5-10
keV band. This indicates a very low accretion rate, in agreement with
the disk instability model for long period systems. However, such a
model predicts stable disks at somewhat long periods. The
discovery of this system outburst, along with similarities to the long
period system SDSS~J080710+485259 with a comparably long, weak outburst, indicates
that these enhanced mass-transfer events may be more common in
long period AM~CVns. A larger sample would be needed to determine empirically at what period, 
if any, the disk instability stops functioning entirely. 
Finally, we identified an infrared excess in the quiescence spectrum
attributable to the donor. This makes SDSS~J113732+405458 the second
AM~CVn to have a directly detected donor.

\end{abstract}




\section{Introduction}
\label{intro}

Binaries with orbital periods ($P_{orb}$) shorter than 70 min \citep[e.g.][ and references therein]{2020Green} in which the primary star is a white dwarf (WD) accreting from a He-rich secondary star are known as AM CVns. After reaching the period minimum \citep[e.g.][]{1985Tutukov}, which depends on the formation channel \citep[see][ for a review]{2010Solheim}, 
these ultracompact white dwarf binaries evolve towards longer orbital periods, decreasing their mass-transfer rate ($\dot M_{tr}$) as their orbit widens 
\citep[e.g.][]{2003deloye,2007Deloye}. Under the disk instability model \citep[DIM, e.g.][]{1983Smak,2008Lasota,2015cannizzo,2019cannizzo} commonly used to explain the outburst activity of AM~CVns, the value of $\dot M_{tr}$ determines the stability of the accretion disk and then the presence of accretion outbursts. In the DIM, a disk is cold and stable if the $\dot M_{tr}$ is everywhere lower than a critical value of the mass accretion rate. The latter depends mainly on parameters such as the mass and size of the accretor, as well as the metallicity of the disk \citep[][]{2008Lasota}. From these conditions it results that long period AM~CVns should have a cold and stable disk (because they have very low $\dot M_{tr}$ values) and therefore show no outbursts. The period limit between outbursting sources and those with cold and stable disks is nonetheless quite uncertain \citep[e.g.][]{kotko2012}.

Using data from the Sloan Digital Sky Survey (SDSS), \citet{2014carter} identified He emission lines and a lack of H, the signatures of AM~CVn stars, in the spectrum of SDSS~J113732+405458 (hereafter, SDSS 1137). The authors also determined a period of $59.6\pm2.7$~min using the radial velocity variations of the emission lines and reported a SDSS-$g$ magnitude of 19~mags during quiescence. Thanks to the recent development of multi-band sky surveys
such as the Zwicky Transient Facility \citep[ZTF,][]{2019masci}, the long term behavior of SDSS~1137 has been monitored. 
Located at RA=11:37:32.32, Dec=+40:54:58.3, the first detected outburst of SDSS~1137 was recorded by ZTF in 2018. In the next sections we present the analysis of the event at different bands and discuss the implications of our results in the context of the DIM.\looseness=-10

\begin{table*}
	\centering
	\caption{X-ray, ultraviolet and optical measurements of SDSS~1137 during quiescence. Other optical measurements have been reported in \citet{2014carter}. Errors in luminosity include error in distance and are given at $1\sigma$, except for those in X-rays, which are given at the $90\%$ confidence level. Magnitudes are not deredened.}
	\label{tab:uvot}
	\begin{tabular}{l |c |c |c | c| c| r}
		\hline
		Instrument & Band & $\lambda_{cen}$ & Exposure & Magnitude & Flux & Luminosity\\
		           &      & (\AA)           &(s)       & (Vega)    & $\times 10^{-14}$ erg/s/cm$^{2}$ & $\times10^{30}$ erg/s \\
		\hline
		Swift/UVOT & $UVW2$ & 1 928 & 17 695 & $18.62\pm0.08$ & $36.8\pm 2.9$ &  $1.92\pm0.30$\\ 
		'' & $U$     & 3 465  & 12 420   & $18.08\pm0.05$ & $71.7\pm3.4$ & $3.75\pm0.55$\\  
		'' & $B$     & 4 392  & 8 290    &$19.18 \pm0.04$ & $60.6\pm2.6$  & $3.17\pm0.46$\\  
		'' & $V$     & 5 468  & 19 150   &$19.29\pm0.10$  & $39.2\pm3.7$  & $2.05\pm0.34$\\ 
		OM & $U$     & 3 472  & 3 635    &$17.94\pm0.03$  &  $83.1\pm2.2$   & $4.34\pm0.57$\\  
		Swift/XRT & 0.3-10 keV & - & 25 000 & - &  $5.2^{+2.8}_{-1.9}$  & $0.27^{+0.17}_{-0.13}$\\
		EPIC-pn   & 0.5-10 keV & - & 11 000 & - &  $5.5^{+1.4}_{-3.4}$  & $0.29^{+0.10}_{-0.21}$\\
		\hline
	\end{tabular}
\end{table*}

\section{OBSERVATIONS AND DATA ANALYSIS}
\label{observations}

In this work we have analyzed data ranging from the X-rays to the near Infrared (NIR) to study the behavior of SDSS~1137 during outburst and quiescence.

\subsection{X-ray and UV data}

A data set of X-ray and $U$ observations was obtained under \textit{XMM-Newton} program ID $080027$ (PI Maccarone) on 2017-12-21.
An exposure time of 11~ks was obtained in X-rays with EPIC-pn and 3.6~ks were gathered with OM-$U$ in imaging mode. 
SDSS~1137 was also detected multiple times within the field of view of the Neil Gehrels Swift Observatory (\textit{Swift}) with the XRT and UVOT detectors. The first observation was obtained on 2005-02-17 and the last one on 2005-03-21. A total of 25~ks were obtained with XRT and 57.5~ks with UVOT in the filters $UVW2$, $U$ and $V$ (see Table \ref{tab:uvot} for more details).

For the X-ray spectral analysis we used the standard products for EPIC-pn from the \textit{XMM-Newton} pipeline processing system\footnote{https://www.cosmos.esa.int/web/xmm-newton/sas-threads}. The OM-$U$ magnitude of SDSS~1137 was calibrated by converting the count rate obtained from the standard products to the Vega and AB systems\footnote{https://www.cosmos.esa.int/web/xmm-newton/sas-watchout-uvflux}. The Swift XRT data were first reprocessed using \textsc{xrtpipeline} and analyzed following the corresponding standard threads\footnote{https://www.swift.ac.uk/analysis/xrt/}. Fitting of the spectra of the XMM-Newton and XRT data was performed using XSPEC v12.11.0 \citep{199arnaud}. We combined the longest XRT observations in which the object was detected and performed a fit. For the EPIC-pn and the XRT data the best fit was an absorbed power-law model (TBabs*pegpwrlw). We set the value of the neutral H column (N$_H$) to the Galactic value ($1.90\times10^{20}$~cm$^{-2}$) and binned the data to 5 and 20 counts per bin for XRT and EPIC-pn, respectively. Given the small number of counts in XRT we used C-statistics for the fit, while for EPIC-pn we used Chi-squared. 

For the UVOT photometric analysis we considered a circular region of $5''$ around the source and estimated the sky background using a region near the source with a radius of $15''$. We then followed the corresponding UVOT threads\footnote{https://www.swift.ac.uk/analysis/uvot/}. 
We created light curves with the UVOT data, but no outbursts were identified in the period covered by these observations.

\subsection{Optical data}

\begin{figure*}
	\includegraphics[scale=1.2]{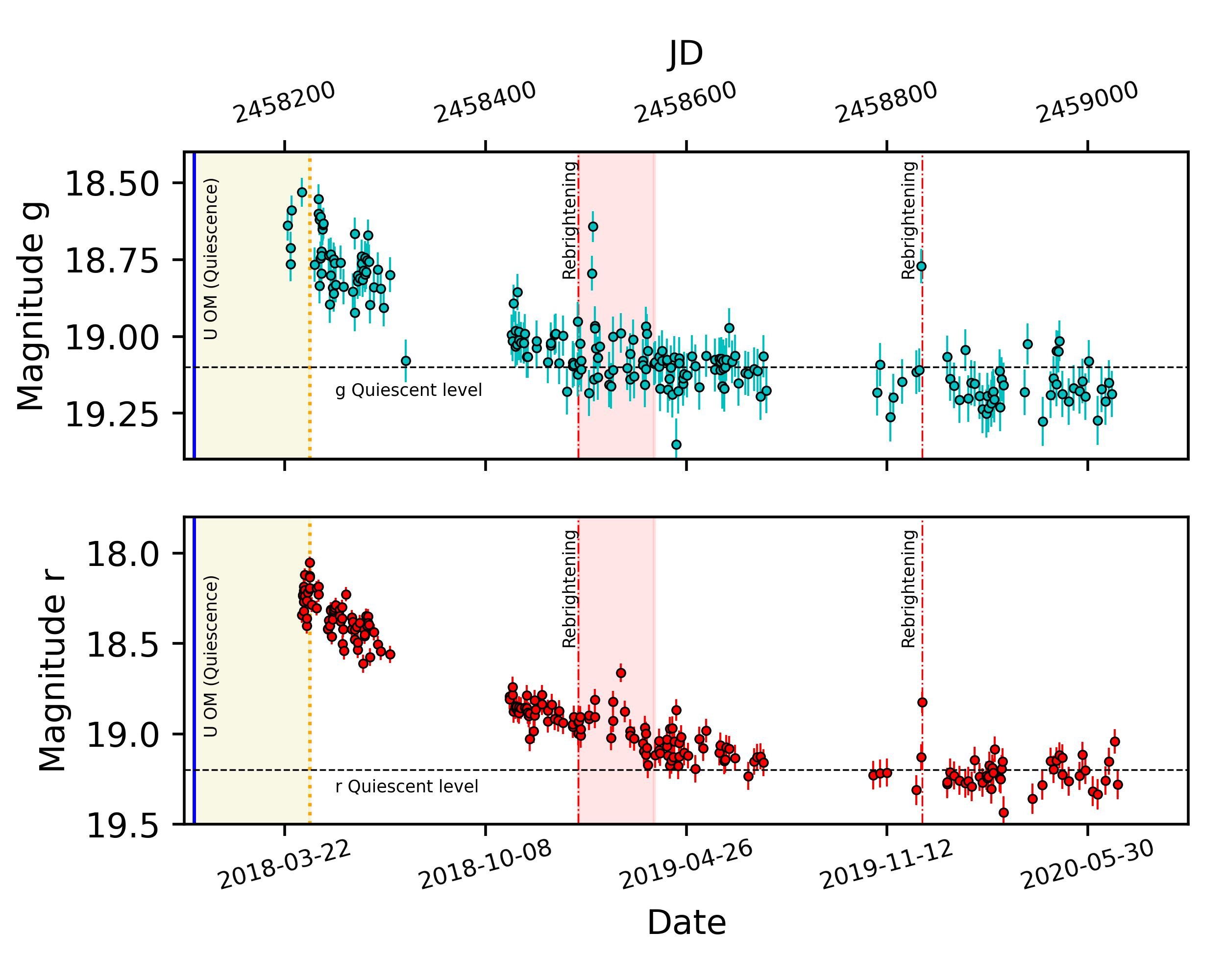}
    \caption{ZTF light curves in the $g$ (top) and $r$ (bottom) bands for SDSS 1137, showing the first detected outburst from this source. The blue vertical line indicates the XMM measurements obtained during quiescence on 2017-12-21. According to ZTF data, the outburst peaked around 2018-04-16 (orange dotted vertical line). The yellow box indicates the possible period between the beginning of the outburst and the peak. 
    We marked the end of the outburst (2019-01-08) when the system showed increasing activity in the $r$ band, which was also observed in the $g$ filter. 
    For comparison, the horizontal dashed lines indicate the quiescence levels in both bands. In $g$ we use the value measured by Pan-STARRS-$g$ (19.1 mags) because it is more similar to ZTF-$g$ than SDSS-$g$. The quiescent level in $r$ is given by both the SDSS and the Pan-STARRS values, which are consistent with each other (19.2 mags). The red region indicates a period of activity in both bands, likely due to rebrightenings. Another rebrightening was detected in December 2019, both in the $g$ and $r$ bands.}
    \label{fig:totalLC}
\end{figure*}

Public ZTF data in the $g$ and $r$ bands were also used in our analysis. The first ZTF measurement was obtained on 2018-03-25 and the last one on 
2020-06-29 (Fig. \ref{fig:totalLC}). There were two gaps in the data coverage due to the source's occultation. We only considered ZTF measurements with good quality flags and excluded points with airmass $ >1.8$ since the differential chromatic refraction that produces color biases dominates above that value \citep{2019masci}.\looseness=-10

Data from the Catalina Sky Survey \citep{2009Drake} show no outbursts between 2005-12-08 and 2013-06-07 and the binary was not found in the DASCH database \citep{2009grindlay}. 
Data from the ASAS \citep{2017kochanek} database show only 3 solid measurements spread in the period 2012-01-11 to 2020-08-01, but several observations performed within minutes from these points just provided upper limits and do not support the measurement values, suggesting that the 3 measurements are likely due to instrumental fluctuations. Available Pan-STARRS data \citep{2016panstarrs} did not reveal any outbursts. We also used data from the AllWISE archive obtained in 2010 during the cryogenic phase \citep{2010wright} in the $W1$ and $W2$ bands where the object was detected. Only good measurements were taken into account.

To compare the outburst behaviour of long period AM~CVns, we have also collected photometric data for the systems SDSS~J080710+485259 (hereafter SDSS~0807, $P_{orb}\approx53$ min) and SDSS~J141118+481257 (SDSS~1411, $P_{orb}\approx46$ min). For the superoutburst of SDSS~0807 \citep{2020RS08} we used ZTF data in the $g$ and $r$ bands, while for the superoutburst of SDSS~1411 we collected AAVSO data in $V$ and $I$ due to the superior coverage compared to ZTF for that source \citep{2019RS}.
We also determined the color evolution of these sources by pairing observations obtained in the above mentioned filters. For SDSS~1137 and SDSS~0807 we combined data points obtained within 1 day. For SDSS~1411 we have paired data points taken within 1.2 hrs from each other in the $V$ and $I$ bands, and later averaged the pairs in bins of 0.5 days given the much shorter duration of the superoutburst and to reduce inhomogeneities in the data acquisition. We have only focused on the second part of the event which was recorded between 2018-05-31 and 2018-06-11. The very first part of the superoutburst of SDSS~1411 has no good coverage in both bands and so it has been excluded.

\begin{figure}
	\includegraphics[width=1\columnwidth, trim=0cm 1cm 0cm 0cm]{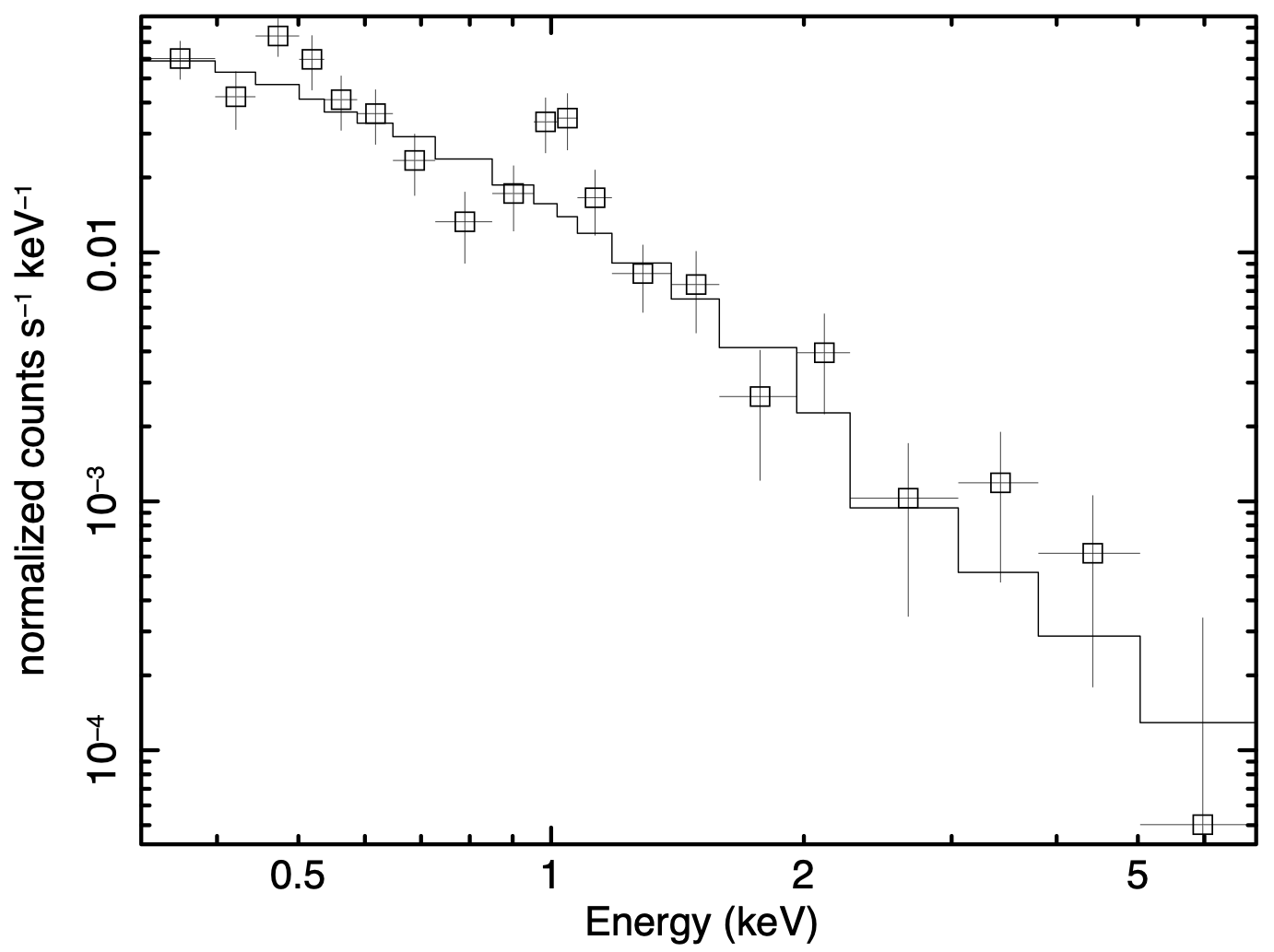}
    \caption{Quiescent X-ray spectrum of SDSS~1137 using the XMM data. The spectrum is better fit with an absorbed power-law model with a photon index $\Gamma=2.5\pm 0.1$. For clarity purposes the image just shows the spectrum in the $0.5-7$~keV energy range but the fit was performed in the $0.5-10$~keV interval.
    }
    \label{fig:spectrum}
\end{figure}

\begin{figure}
	\includegraphics[width=\columnwidth]{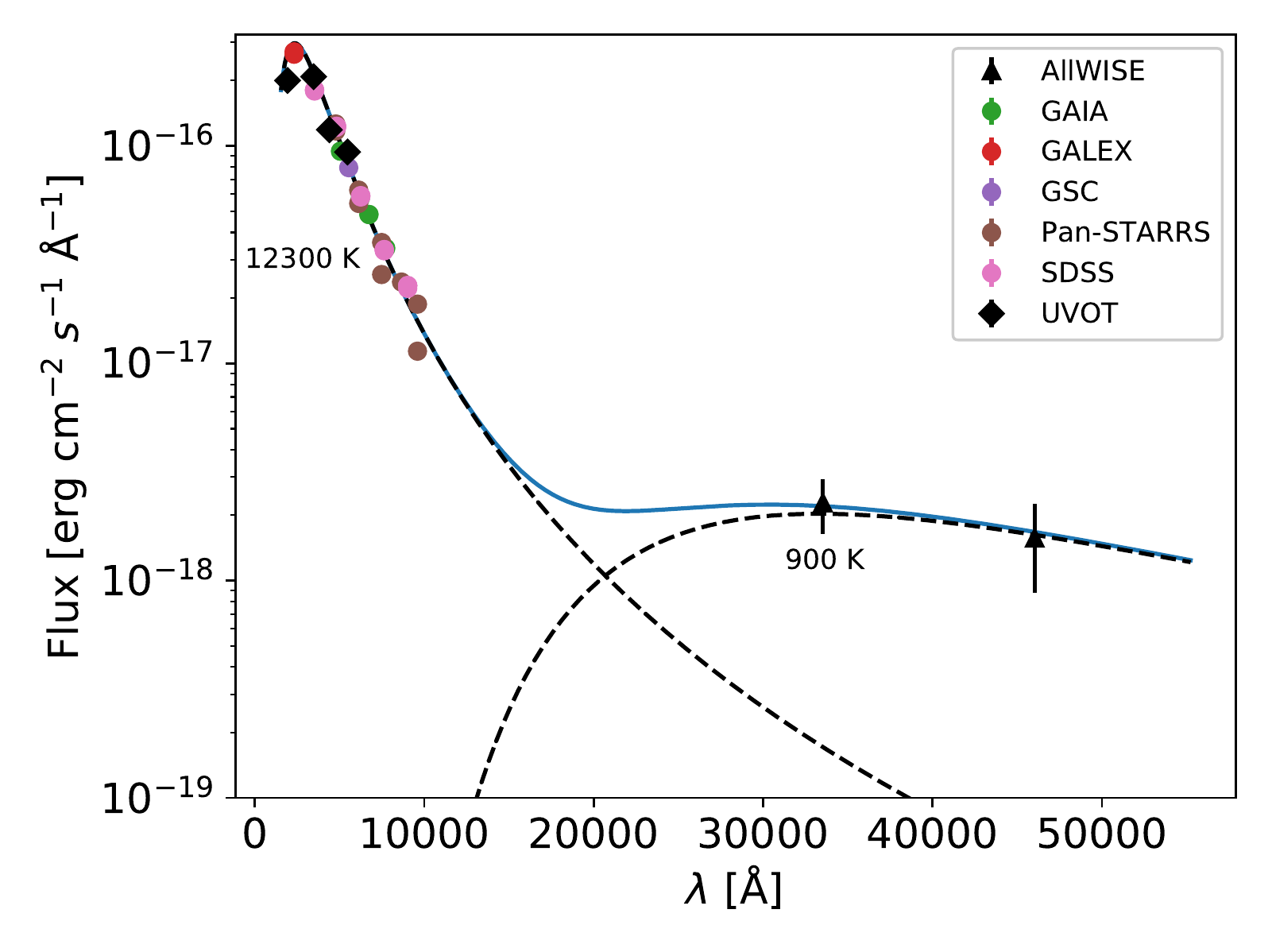}
    \caption{Spectral energy distribution (SED) for SDSS~1137 in quiescence using observations from the NUV to the NIR. We fitted two black bodies to the data (dashed lines), the temperatures of which are indicated in the figure. The sum is indicated as a blue solid line. From the fit it is evident that an excess is present in the NIR flux with respect to the hottest black-body and it is likely coming from the donor star. While the obtained temperature of the hottest black-body is consistent with spectral analysis \citep{2014carter} and the flux is largely dominated by the WD accretor, the temperature of the cold black-body is just indicative, as there are large uncertainties in the donor's size. Nonetheless, the NIR excess is obvious. Data for the fit was deredened.}
    \label{fig:SED}
\end{figure}

\begin{figure}
	\includegraphics[width=1.01\columnwidth, trim=0cm 0cm 0cm 0cm]{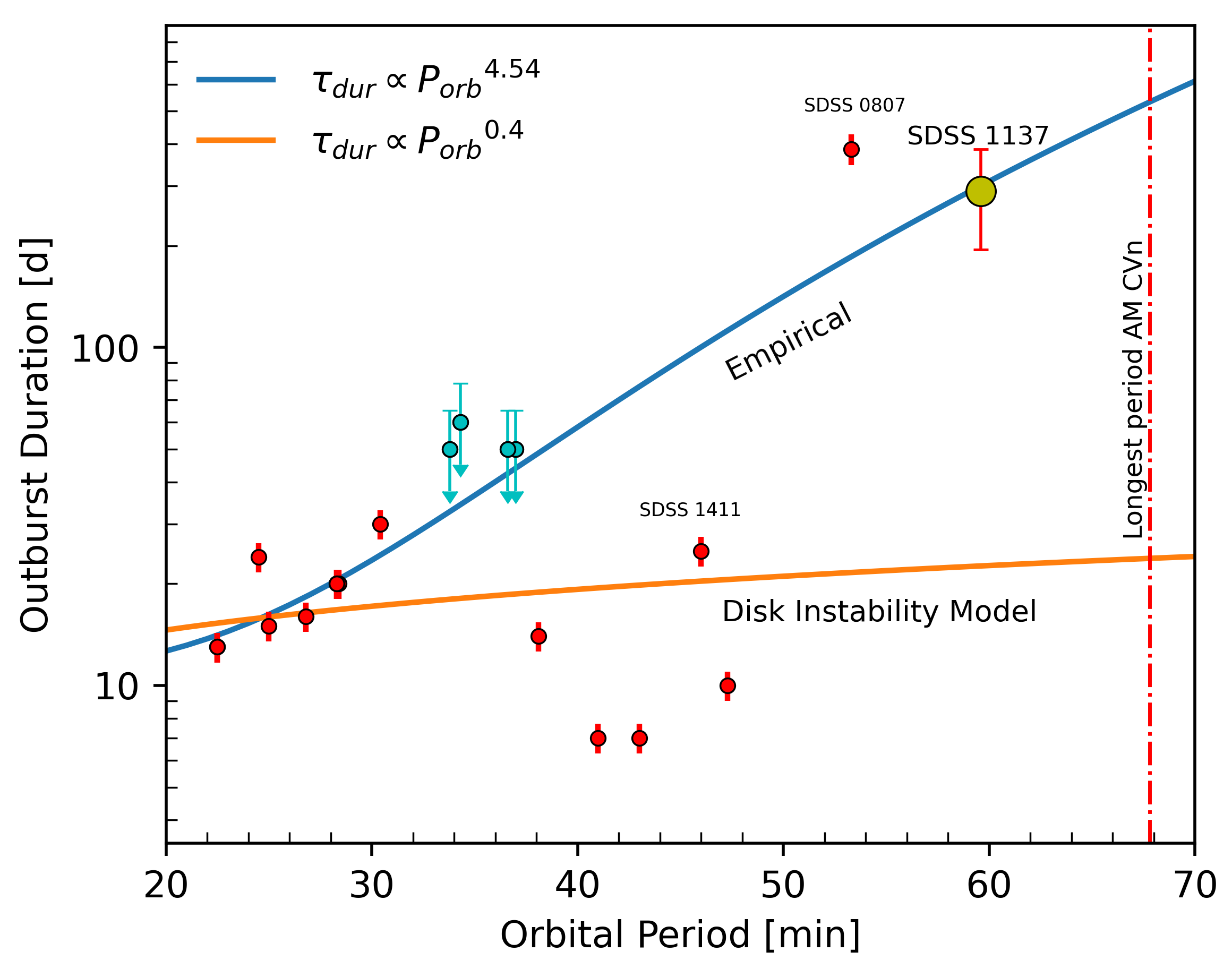}
    \caption{Orbital period ($P_{orb}$) vs outburst duration ($\tau_{dur}$) for AM~CVns. The orange line indicates the relation obtained from the DIM \citep{2019cannizzo} and the blue line the empirical relation \citep{2015Levitan} derived for systems with $P_{orb}$ up to 37 min. Upper limits are marked in cyan color.    
    We superimpose the results for SDSS~1137 from this paper, for SDSS~0807 \citep{2020RS08} and for SDSS~1411 \citep{2019RS}. The first two systems clearly exceed the expected $\tau_{dur}$ from the DIM by several times suggesting the presence of additional (or even different) mechanisms in the origin of the outbursts, while SDSS~1411 still fits the predictions from the model. Note that the duration of the outburst in SDSS~1137 is an upper limit, where the yellow circle denotes the duration between the first ZTF measurement and  the chosen date for the end of the outburst (See Sec. \ref{discussion}).
    The dashed line indicates the AM~CVn with the longest $P_{orb}$ identified so far (SDSS~J1505+0659, $P_{orb}=67.8$ min) as reported by \citet{2020Green} and which up to date has not shown outbursts. 
    Other data in the plot were taken from \citet{2019cannizzo} and references therein. A $10\%$ error for the outburst duration for all systems that do not have an upper limit was considered, as done by \citet{2015Levitan}.
    }
    \label{fig:dur_per}
\end{figure}

\begin{figure*}
    \begin{minipage}[t]{.49\textwidth}
        \centering
        \includegraphics[width=\textwidth]{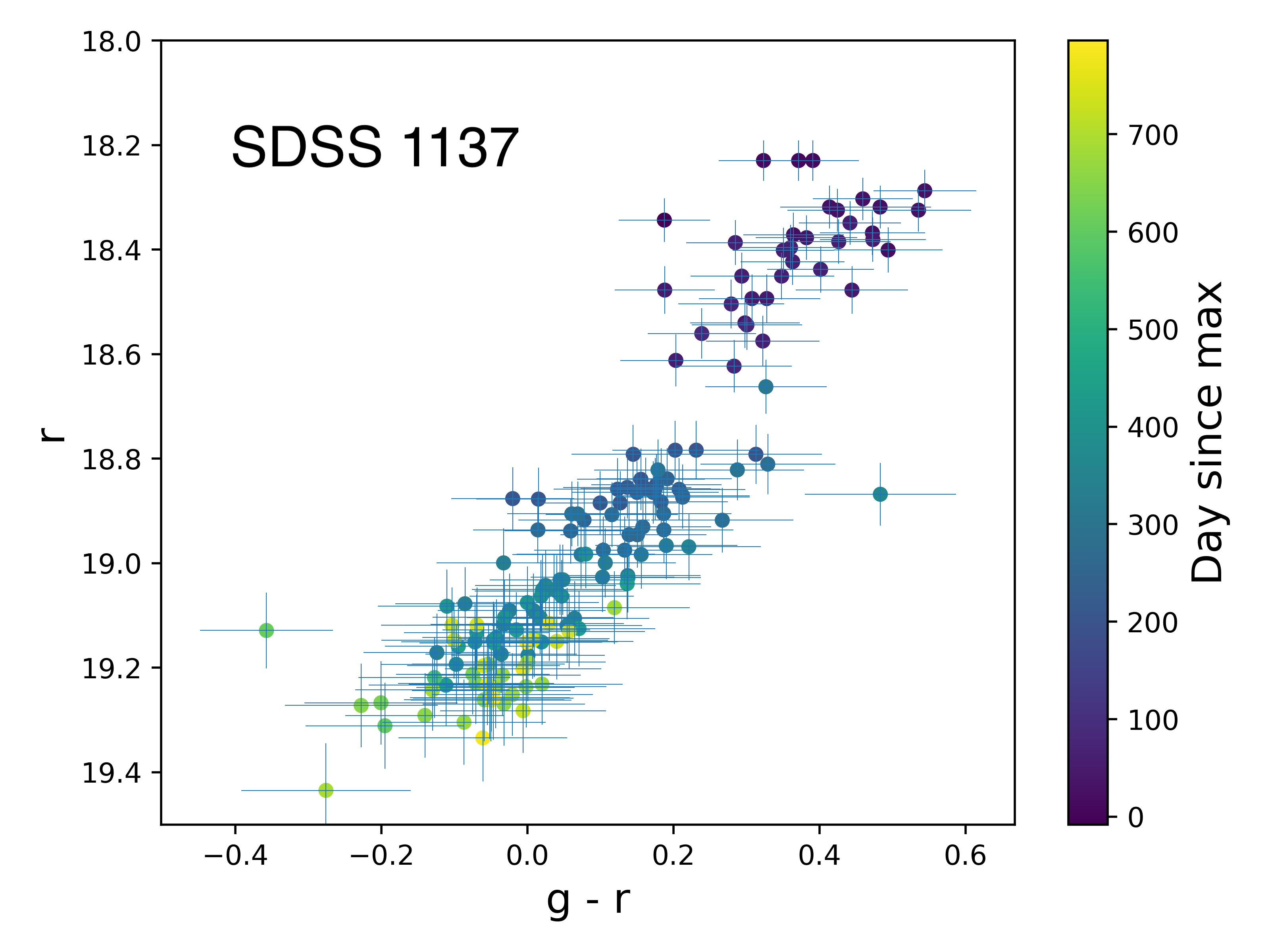}
    \end{minipage}
    \hfill
    \begin{minipage}[t]{.49\textwidth}
        \centering
        \includegraphics[width=\textwidth]{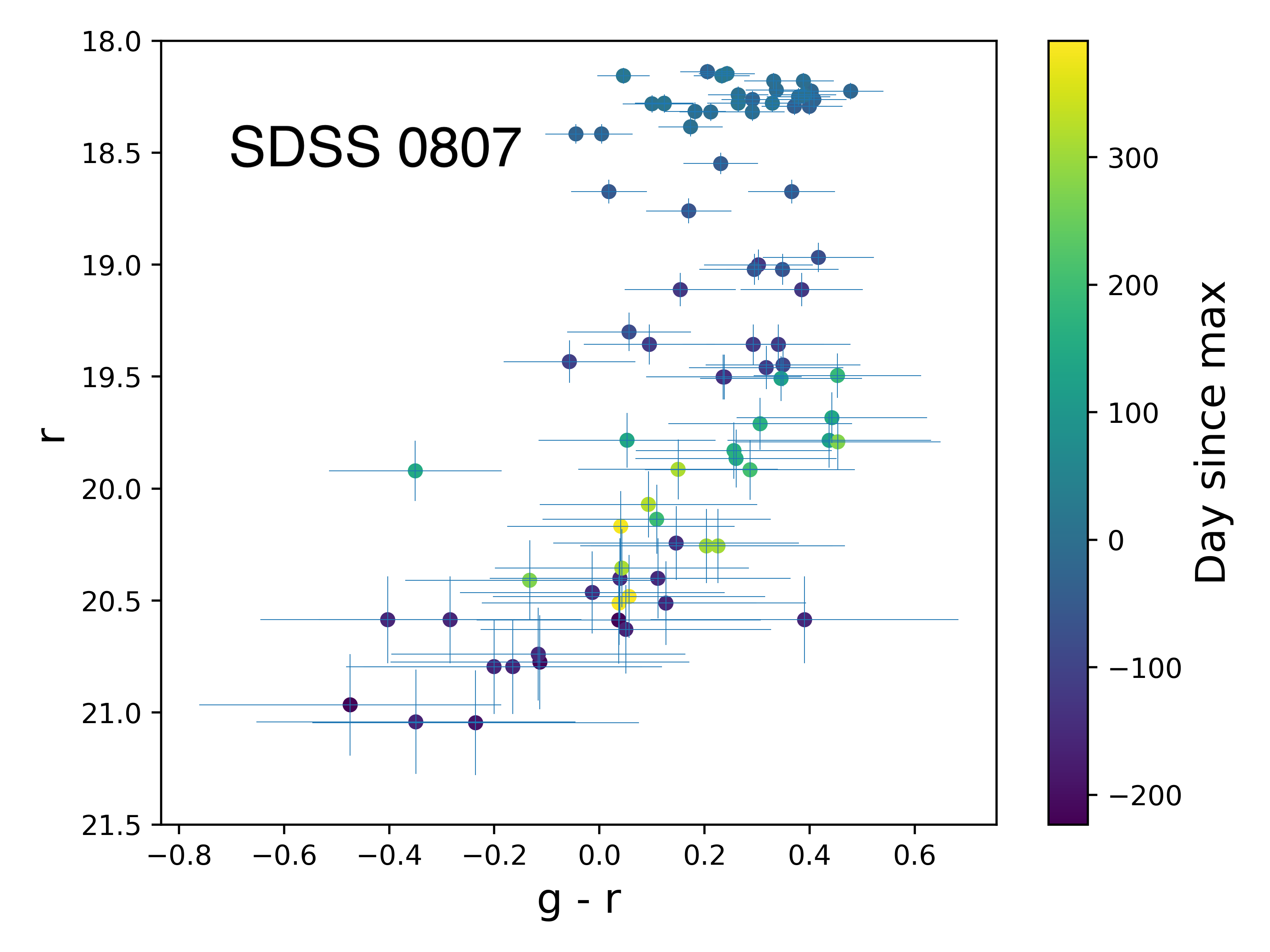}
    \end{minipage}
    \hfill
    \caption{Color variations for the AM~CVns SDSS~1137 ($P_{orb}\approx60$ min, left) and SDSS~0807  ($P_{orb}\approx53$ min, right) during outburst. During quiescence the binaries are blue and dominated by the emission from the primary WD. As the systems are in the rise phase they become redder and brighter because the accretion disk becomes hotter. In the case of SDSS~0807 the binary turns bluer near the peak, suggesting that the accretion disk is now the dominant source and that likely a disk instability was triggered or the disk heated enough due to the mass enhancement. However, the same behavior was not observed in SDSS~1137, suggesting that the accretion disk in that system never reached the temperatures to ionize He. In both systems an enhanced mass-transfer is the likely origin of the outbursts.}
    \label{fig:hysteresis1137}
\end{figure*}

\begin{figure}
	\includegraphics[width=\columnwidth]{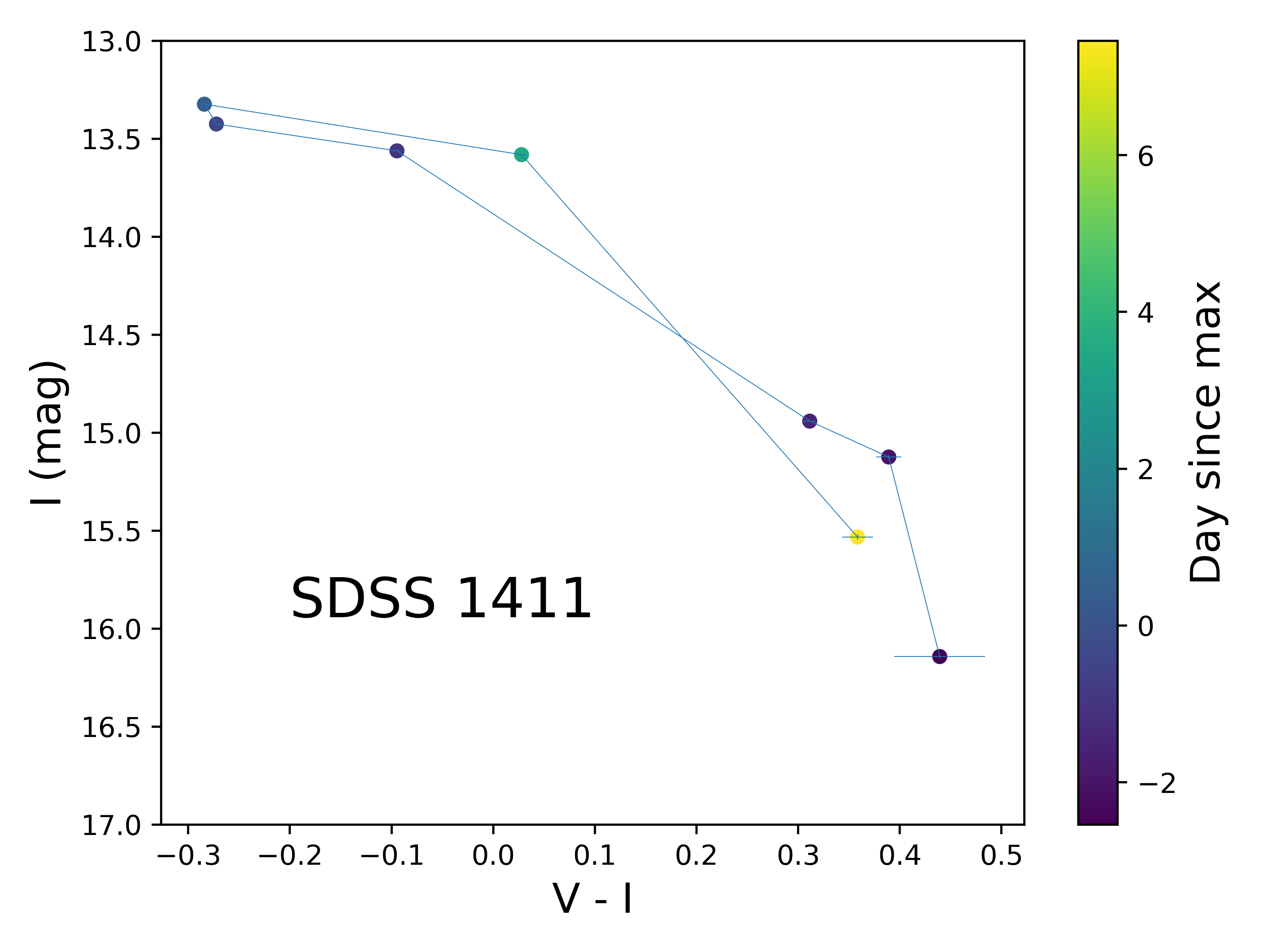}
    \caption{Color variations for the AM~CVn SDSS~1411 ($P_{orb}\approx46$ min) during superoutburst. The binary becomes bluer during the rise phase. During peak it is bluest and brightest and during decay it becomes redder again. Note that despite in this case the analysis has been done using the $I$ band instead than $r$, the binary follows a completely different pattern  when compared to longer period AM~CVns (Fig. \ref{fig:hysteresis1137}). The direction followed by SDSS~1411 is similar to that of DNe outbursts \citep{2020hameury} and it is well explained under the DIM. Data points correspond to observations in $V$ and $I$ obtained by the AAVSO observers.  When several observations were obtained very close to each other, we have paired them and averaged them (with their respective errors) in bins of 0.5 days to remove the inhomogeneities of the data.
    }
    \label{fig:hysteresis1411}
\end{figure}

\section{RESULTS}
\label{sec:results} 

\subsection{The multi-wavelength light curve of SDSS 1137}

The X-ray spectral fit of the EPIC-pn data showed that the spectrum of the binary could be described by an absorbed power-law with a photon index $\Gamma=2.5\pm 0.1$ and had a count rate in the $0.5-10$ keV band of $25.5\pm2.19\times10^{-3}$ counts s$^{-1}$ (Fig. \ref{fig:spectrum}). This corresponds to a flux  $f_X=5.5^{+1.4}_{-3.4}\times10^{-14}$ erg s$^{-1}$ cm$^{-2}$ equivalent to a luminosity $L_X = 2.9^{+1.0}_{-2.1}\times10^{29}$ erg s$^{-1}$ at the distance of $209 \pm 11$~pc \citep{2020gaia}. These values are consistent with the ones obtained with XRT in the range $0.3-10$~keV: $\Gamma=2.4^{+0.9}_{-0.6}$ and  $f_X = 5.2^{+2.8}_{-1.9}\times10^{-14}$ erg s$^{-1}$ cm$^{-2}$. The larger uncertainty in the Swift values is due to the smaller number of counts detected, since XRT is much less sensitive than EPIC-pn. All errors are provided at $90\%$ confidence level.

From the OM-$U$ data we obtained a magnitude of $17.94\pm0.03$ mag in the Vega system which is consistent with that from UVOT-$U$ ($18.08\pm 0.05$), suggesting that on 2017-12-21 SDSS~1137 was still in quiescence.
The consistency in the X-ray values of both telescopes also supports a scenario in which both measurements were obtained in quiescence. 
However, the long term lightcurve (Fig. \ref{fig:totalLC}) with the ZTF data shows that on 2018-03-25 the binary was above its Pan-STARRS and SDSS-$r$ quiescent level of 19.2~mags \citep{2014carter}, revealing an outburst\footnote{Note that in this particular case we have not named the event a superoutburst despite its long duration. Reasons will be explained in Sec. \ref{discussion}.} which reached its peak around 2018-04-16\footnote{After the submission of this manuscript, we became aware of the existence of additional data that may suggest that the outburst reached its maximum much earlier than here described. The findings reported in this document are based solely on the available data from the databases mentioned in Sec. \ref{observations}.}.
ASAS data from 2017-12-21 to 2018-03-29 indicate that SDSS~1137 was never detected above 18~mags\footnote{The  $g$ ASAS data point of 18.09~mags on 2017-12-26 is considered an upper limit for the reasons mentioned in Sec.\ref{observations}.}, thus imposing an upper limit of 1~mag on the amplitude of the outburst. The duration of the event is difficult to determine given the peculiar behavior in the $g$ and $r$ bands: SDSS~1137 seems to have reached quiescence much earlier in $g$ than in $r$ (this behavior will be discussed in the next section). Consequently, we placed the end of the outburst when there was a period of increasing activity in the $r$ band (around 2019-01-08), which was also observed in the $g$ filter, suggesting that likely there were additional smaller rebrightenings in that period. Since they occurred during the decay of the outburst, it is possible that they might be echo outbursts. Afterwards there seems to be another period of increasing brightness starting on 2019-12-17 more visible in the $r$ band, where consecutive measurements follow a trend. The fact that the increase in activity is observed in both filters, together with the relative low air-mass values, good photometric flags, as well as small amplitude show that that rebrightening is real and likely due to a smaller outburst\footnote{It is unclear whether that rebrightening could be an echo outburst or a normal outburst.}.

\subsection{Spectral energy distribution}

We plotted the spectral energy distribution (SED) of the binary and performed a black-body fit of the data points in order to determine its temperature and normalization. We combined our UVOT measurements, the available information from the SDSS, Gaia \citep{2018gaia}, Pan-STARRS, as well as AllWISE data of SDSS~1137, all obtained during quiescence (i.e. before December 2017). Data for the fit was deredened using $E(B-V)=0.018$ \citep{1998schlegel, 2019greenmap} and $R_V=3.1$.

From our SED analysis using observations from the NUV to the NIR (Fig. \ref{fig:SED}) it is clear that a single black-body can not reproduce all observations. The AllWISE data show that there is a significant NIR excess. Based on these results we have then added another black-body component. 
We did not fix any parameter and obtained temperatures of $12.3\pm 0.5 \times 10^3$ K for the hottest component and $9^{+26}_{-8} \times 10^2$ K for the coldest one\footnote{The errors on the temperature of the second black-body are large, as are the ones on its normalization, because we only have two data points from the AllWISE survey for the fit.}.

\section{DISCUSSION}
\label{discussion}

The detected outburst activity in SDSS~1137 is an unprecedented event considering its large orbital period. In fact, with P$_{orb}=59.6$~min, it is the longest period outbursting AM~CVn detected so far. Since the distance to the binary is known ($209 \pm 11$~pc), we have derived an absolute UVOT-$V$ magnitude $M_V=12.69$ mag during quiescence. In the same state and at its orbital period the optical emission is mostly coming from the accreting WD because the disk is cold. 
This result is consistent with what is expected for a massive WD accretor with $M=1.05M_{\sun} $ \citep{2006bil}. A massive accretor is also supported by the small radius obtained from the normalization of the hottest black-body in our SED analysis. The expected $T_{eff}$ ($\sim 12\ 000$~K) is also very well consistent with the one we have obtained from our SED fit (Fig. \ref{fig:SED}). Assuming a massive WD ($1.05 M_{\sun} $), \cite{2006bil} predict that the average accretion rate for a $P_{orb}=60$ min system is around $3\times10^{-12}M_{\sun}/yr$. For SDSS~1137, if we consider the obtained X-ray luminosity and even (unrealistically) all the UV emission as coming from the accretion process\footnote{Considering our SED fitting the obtained UV luminosity is likely coming mostly from the accreting WD, but for the mass accretion calculation using $L_{acc}=GM\dot M/R$ we have considered the observed UV luminosity in order to obtain a better estimate than just using the X-rays.}, we determine a mass-accretion rate during quiescence of $\sim1.4\times10^{-13}M_{\sun}/yr$, for $M=1.05M_{\sun}$ and a radius of $0.008R_{\sun}$ (similar to Sirius B). This rate is in fact smaller than the one of the critical mass-accretion value for a similar mass WD and the viscosity parameter $\alpha=0.1$ \citep{2008Lasota}, which is $\dot M_{crit}\approx2.2\times10^{-13}M_{\sun}/yr$, in agreement with expectations for the DIM that the disk should be stable.
Furthermore, our determined quiescent $\dot M$ value is an order of magnitude smaller than the one estimated by \cite{2018ramsay} which is $8.1\pm 3.9 \times10^{-12}M_{\sun}/yr$ (determined during quiescence by fitting a model consisting of a WD plus a steady-state accretion
disk model to the SED of SDSS~1137). The large difference may be due to the parameters considered. \cite{2018ramsay} assumed an accretor with $M=0.8\pm0.1 M_{\sun}$ and an inclination value of $\cos i =0.5$, but, more importantly, these authors varied the size of the accretion disk to reach a good fit. They determined the inner and outer radii to be $R_{in}=0.012\ R_{\sun}$
and $R_{out}=0.018\ R_{\sun}$, respectively. Therefore, the discrepancy in $\dot M$ could also be due to the real size of the disk. 
Unfortunately, during outburst we cannot determine the value of the mass accretion rate since we do not have X-ray nor UV coverage, and using the optical values is not appropriate.

It is interesting to note that the (one order of magnitude) smaller observed X-ray and UV luminosities of SDSS~1137 during quiescence (see Table \ref{tab:uvot}), when compared to other long period systems such as GP~Com \citep{2005ramsay} and SDSS~1411 \citep{2019RS}, also support a very low accretion rate, as expected for its evolutionary stage. From our observations of SDSS~1137 we also see that the binary mostly emits in blue bands as occurs in other disk accreting AM~CVns. The X-ray emission is an order of magnitude smaller than the UV one, and the X-ray spectrum is also soft (consistent with the one of SDSS~1411 within errors). This suggests that the X-ray component is less important in AM~CVns with disks than the UV component. 

The main uncertainty in the lightcurve of Fig. \ref{fig:totalLC} 
is the duration of the outburst. This is due to the poor photometric coverage at early epochs (between 2017-12-21 and 2018-03-25), as well as the instrumental fluctuations and relatively large photometric uncertainties of the ZTF data, especially in the $g$ band. However, an upper limit can be established considering the observation date of the XMM-Newton data (2017-12-21) and the beginning of the first rebrightening, which then imply an outburst duration of $<380$ days. For comparison purposes we have plotted that value in the outburst duration vs orbital period relation for AM~CVns (Fig. \ref{fig:dur_per}). This duration is comparable to that of the recently reported value for SDSS~0807 \citep{2020RS08}, for which the first detected superoutburst lasted 390 days. It is clear that these 2 systems do not follow expectations from the DIM. They exceed by several times their expected duration, suggesting that there are additional mechanisms at work causing the outbursts of (long period) AM~CVn binaries.

\subsection{Color evolution during outbursts of AM~CVns: evidence of enhanced mass-transfer}

In order to investigate the mechanism(s) that drive outbursts in long period AM~CVns, we performed an analysis of the color evolution during outburst of the 3 AM~CVns SDSS~1137, SDSS~0807 and SDSS~1411, which are long period outbursting systems for which we have coverage in more than one band. 
In Fig. \ref{fig:hysteresis1137} and \ref{fig:hysteresis1411} we show the color magnitude diagram results for each binary at different stages during the outburst. One can see that SDSS~1137 and SDSS~0807 follow a completely different pattern in their evolution compared to the one of SDSS~1411. The latter seems to be consistent with the path followed by cataclysmic variables of the type dwarf novae
\citep[DNe,][]{1980bailey, 2020hameury} and it can be well explained under the DIM. 

In the case of SDSS~1137 the color evolution track starts with a red color and large brightness. This is due to the fact that there are no multiple close pairs ($<1$ day) of data in the bands ZTF-$g$ and ZTF-$r$ before maximum (see Fig. \ref{fig:totalLC}). The binary then becomes bluer as it decays from the outburst, reaching its bluest when it is back in quiescence. This is due to the dominant emission of the accreting WD. One should note that despite the large errors in color, it is clear that even during the peak, the binary never reached the temperatures (visible through its colors) necessary to ionize He, the dominant element in the disk. Though the disk became hotter and the binary brighter than in quiescence, the disk never dominated the emission generated by the other components in the system (the accreting WD and likely the hot spot). \emph{This is an extremely important result as it then suggests that the observed outburst was unlikely related to instabilities in the accretion disk}. Instead, it is likely that other mechanisms such as enhanced mass-transfer from the companion are responsible for such behavior. Under this scenario the additional mass travels through the accretion disk heating it up, but not triggering a major instability. That would explain the long rise time ($20<\tau_{dur} (days)<120$), as the mass travels all the way from the outer edge to the inner parts, while the disk remains in the low $\alpha$, cold state. The mechanisms
causing the mass-transfer enhancements are unknown, but they could be similar to those hypothesized for some anomalous cataclysmic variables of the type Z Cam  \citep{2014hameury}, for example the donor's magnetic activity. However, since in the case of long $P_{orb}$ AM~CVns the donor is likely a WD, the magnetic fields would need to be large \citep[e.g.][]{Moussa2020}. 
This is probably unrealistic considering the expected masses for AM~CVn donors, but further modeling and observations are needed to investigate that. Mass-transfer changes could also be due to spots on the donor \citep{1976landi,2014hameury,2015Kilic}, in which case magnetic fields would be involved as well. Additionally, it has been shown that mechanisms such as the irradiation of the companion in AM~CVns \citep{1995Warner,1997hameury,2007Deloye,kotko2012,2015warner} and on the WD donors in ultra-compact X-ray binaries \citep{2017liu} plays an important role, severely influencing the evolution of these systems by triggering high mass-transfer rates, extending the donor's adiabatic expansion phase \citep{2007Deloye} to longer periods and likely affecting their period distribution across their  evolution.
Also, irradiation has been shown to be an important ingredient in the modeling of outbursts \citep[e.g.][]{kotko2012}. Additional mechanisms that may lead to fluctuations in the donor's mass-transfer (e.g. pulsations) remain to be investigated.

For the binary SDSS~0807 we observe a somewhat similar cycle, but with a few differences. During quiescence the binary is blue, which is due to the dominating emission from the primary WD. During the rise phase of the superoutburst the binary becomes redder and brighter, again due to the accretion disk becoming hotter. However, unlike SDSS~1137, this system turns to blue colors near the peak of the superoutburst suggesting the disk emission became dominant at this stage. This may indicate that eventually an instability was triggered in the disk, or that the mass flow across the disk, induced by the enhanced transfer, substantially heated up the disk. As the system returned back to quiescence the system became less bright. However, based on data up to mid-2020 \citep[][]{2020RS08}, the binary had not reached full quiescence, and the colors remained redder when compared to colors before the outburst. 

Contrary to the previously described cases, the AM~CVn SDSS~1411 clearly becomes bluer as it moves on the rising phase of the outburst, and it is bluest and brightest during the peak. Afterwards the binary becomes redder and fainter reaching colors similar to those before the superoutburst. This behavior is very well consistent with the one for DNe outburst \citep{2020hameury}, which explains why SDSS~1411 agrees with expectations from the DIM for He dominated disks \citep[Fig. \ref{fig:dur_per} and ][]{2019cannizzo}.

It is also worth to note the low amplitude of the outburst in SDSS~1137. It has been previously shown that the empirical relation between the orbital period and the outburst amplitude derived by \cite{2015Levitan} seems to fail for long period AM~CVn \citep{2019RS,2020RS08}. This is not surprising considering that such relation was derived for systems with orbital periods no longer than 37 min. In the case of SDSS~0807, the low amplitude could be explained (at least partially) as due to inclination effects, but the relatively small radial velocity of SDSS~1137 \citep[$\leq102.47 \pm 19.84$ km/s,][]{2014carter} compared to the one of SDSS~0807 (or even SDSS~1411) argues against a similar explanation. Metallicity has an important effect on the amplitude of outbursts in AM~CVns \citep{kotko2012}, making the amplitude smaller as the metallicity increases. However, while metallicity could have influenced the particularly small amplitude of SDSS~1137, it is more likely that such amplitude is due to the nature of the outburst itself, i.e. not related to disk instabilities as we have discussed above. 

It is interesting to note a trend in the outburst behaviour as a function of the period. The shortest period source in our sample, SDSS~1411 ($\approx 46$~min), shows a disk instability since the beginning of the outburst. As we move to a longer period, SDSS~0807 ($\approx 53$~min), mass-transfer enhancement seems to be the main mechanism at onset, but eventually a disk instability was probably triggered (or at least the disk was substantially heated up by the enhanced mass transfer). Finally, at even longer periods, SDSS~1137 ($\approx 60$~min), the triggering process seems to be still mass-transfer enhancement, but the disk is too cold to ever dominate the emission/lightcurve. Of course, these periods are by no means definitive limits between the different behaviours, but they do seem to indicate a trend.

In order to investigate the period limit (if any) between enhanced mass-transfer driven outbursts and those due to disk instabilities, more long period AM~CVns in outburst have to be studied. Some authors \citep[e.g.][]{kotko2012} have placed such period limit between 50 and 65 min, with the exact limit depending on the mass of the accreting WD and the metallicity of the disk. For example, for a massive WD (1 M$_{\odot}$) and a pure He disk that limit seems to be around 65 min. Thus, it is possible that if the accretor in SDSS~1137 is a massive one, disk instability outburst can eventually be triggered (additionally to the mass-transfer ones like the one presented in this paper), but likely with an extremely long recurrence time.

\subsection{The companion of SDSS 1137}

From our SED analysis we determined the existence of a NIR excess which could be attributed to the companion star, making this the second direct detection of the donor in a non-direct impacting AM~CVn \citep[the other one is SDSS~J1505+065 recently reported by][]{2020Green}.
The detected NIR excess helps explaining the extreme red colors of SDSS~1137 which make the binary a clear outlier in the $g-r$ vs $u-g$ color-color diagram \citep{2014carter}.
Considering the $W1-W2$ color, the companion would be consistent with a star of spectral type L or T \citep{2013pecaut}. That spectral type is then consistent with the one of SDSS~J1505+065. However, the detection of broad N emission lines in the red parts of the spectrum of SDSS~1137 seems to rule out a highly evolved He star donor \citep{2014carter} and instead, it favors a WD scenario \citep{2010nelemans}. In this paper we are unable to be more specific regarding the type of donor given the few NIR measurements available and the large photometric uncertainties in the two AllWISE data points which also affect the SED fit of Fig. \ref{fig:SED}. In fact, the temperature and normalization of the second black-body fit are only indicative, but they demonstrate a clear NIR excess in the SED. 

\begin{figure}
	\includegraphics[width=\columnwidth]{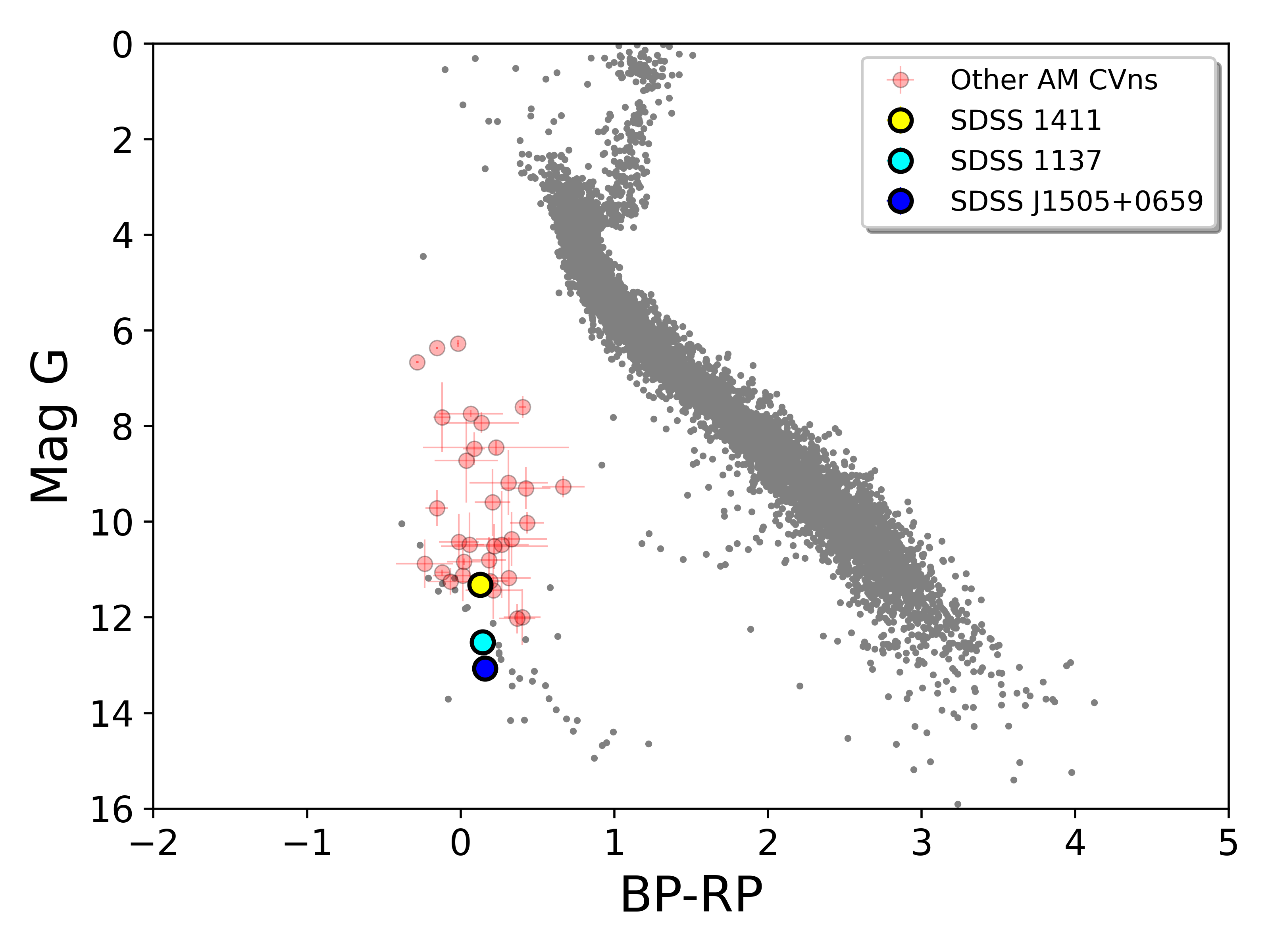}
    \caption{Color-Magnitude Diagram of the $2.5\deg$ area around SDSS~1137 using Gaia EDR3 data \citep{2020gaia}. The AM~CVn systems SDSS~1411, SDSS~1137 and SDSS~J1505+065 lie on the WD cooling sequence, clearly showing the dominant emission of the accreting WD during quiescence. The faintness of the systems is correlated to their orbital period. There is no good parallax information for SDSS~0807 and therefore it is not shown. 
    Gray points correspond to stars with parallax error less than 10\%. Red points correspond to other AM~CVns. No extinction correction have been applied to the stars in this plot.}
    \label{fig:gaia}
\end{figure}

We have further compared the position of SDSS~1137 and SDSS~J1505+065 in a \textit{Gaia} Color-Magnitude Diagram in Fig.\ref{fig:gaia}. Both AM~CVn systems lie on the WD cooling sequence, confirming that the contribution from the companion is really low. Instead, the emission is dominated by the accreting WD with the brightness of the object correlated to their orbital period, i.e. longer orbital period AM~CVns are older and therefore fainter.

Note that given the current observations, we cannot discard the presence of a circumbinary disk, the emission of which peaks in the NIR \citep[e.g.][]{2001spruit,2004dubus}. More data are required in order to investigate each scenario.\\

While writing this manuscript we became aware that \cite{2021Wong} were writing a research note reporting the detection of the same long duration outburst. 

\section*{Acknowledgements}

We thank the anonymous referee for her/his comments which improved this manuscript. We also thank Dr. J.M. Hameury for useful discussion. LERS is supported by an Avadh Bhatia postdoctoral Fellowship at the University of Alberta. This paper is based on observations obtained with the Samuel Oschin 48-inch Telescope at the Palomar Observatory as part of the Zwicky Transient Facility project. ZTF is supported by the National Science Foundation under Grant No. AST-1440341 and a collaboration including Caltech, IPAC, the Weizmann Institute for Science, the Oskar Klein Center at Stockholm University, the University of Maryland, the University of Washington, Deutsches Elektronen-Synchrotron and Humboldt University, Los Alamos National Laboratories, the TANGO Consortium of Taiwan, the University of Wisconsin at Milwaukee, and Lawrence Berkeley National Laboratories. Operations are conducted by COO, IPAC, and UW. This work has made use of data from the European Space Agency (ESA) mission
{\it Gaia} (\url{https://www.cosmos.esa.int/gaia}), processed by the {\it Gaia}
Data Processing and Analysis Consortium (DPAC,
\url{https://www.cosmos.esa.int/web/gaia/dpac/consortium}). Funding for the DPAC
has been provided by national institutions, in particular the institutions
participating in the {\it Gaia} Multilateral Agreement. The authors also acknowledge the SDSS, AAVSO and VizieR data bases for providing part of the data presented in this manuscript.

\section*{Data Availability}

All data used in this document are public and they can be found in the ZTF, XMM-Newton and Swift archives and on the VizieR and AAVSO webpages.



\bibliographystyle{mnras}
\bibliography{biblio} 








\bsp	
\label{lastpage}
\end{document}